# OO Model of the STAR offline production "Event Display" and its implementation based on Qt-ROOT

V.Fine
*BNL, Upton, NY 11973, USA*

J.Lauret
*BNL, Upton, NY 11973, USA*

V.Perevoztchikov
*BNL, Upton, NY 11973, USA*

The paper presents the "Event Display" package for the STAR offline production as a special visualization tool to debug the reconstruction code. This can be achieved if an author of the algorithm / code may build his/her own custom Event Display alone from the base software blocks and re-used some well-designed, easy to learn user-friendly patterns. For STAR offline production Event Display ROOT with Qt lower level interface was chosen as the base tools.

## 1. 3D VISUALIZATION AND HEP TASKS

The paper presents the "Event Display" package for the STAR offline production as a special visualization tool to debug the reconstruction code.

3D visualization has not found for in High Energy Physics research. That essentially represents one or very few events and it can not delivery the information the experiment is designed for. This means the scope of such approach is restricted with:

- Debugging the detector geometry
- Debugging the simulation, reconstruction and data analysis software codes
- Public relations
- Rare event evaluations
- "Control room" like applications.

The list implies the functionality such "Display" application is to be changed as soon as some piece of the instantly evolving code demands the "Event Display" to help finding and recognizing some new problem.

The Object-Oriented approach allows creating the reliable OO model and its C++ implementation that is flexible enough to match the oncoming debugging goals. To be useful the working program must be done quickly otherwise the code will be debugged by other means.

This can be achieved if an author of the algorithm / code may build his/her own custom Event Display alone from the base software blocks and re-used some good and easy to learn user-friendly patterns.

## 2. "STAR" PRODUCTION EVENT DISPLAY

### 2.1. Object-Oriented model of 3D visualization

For STAR offline production "Event Display" ROOT with Qt lower level interface was chosen as the base tools [1]. That allows including the implementations of the

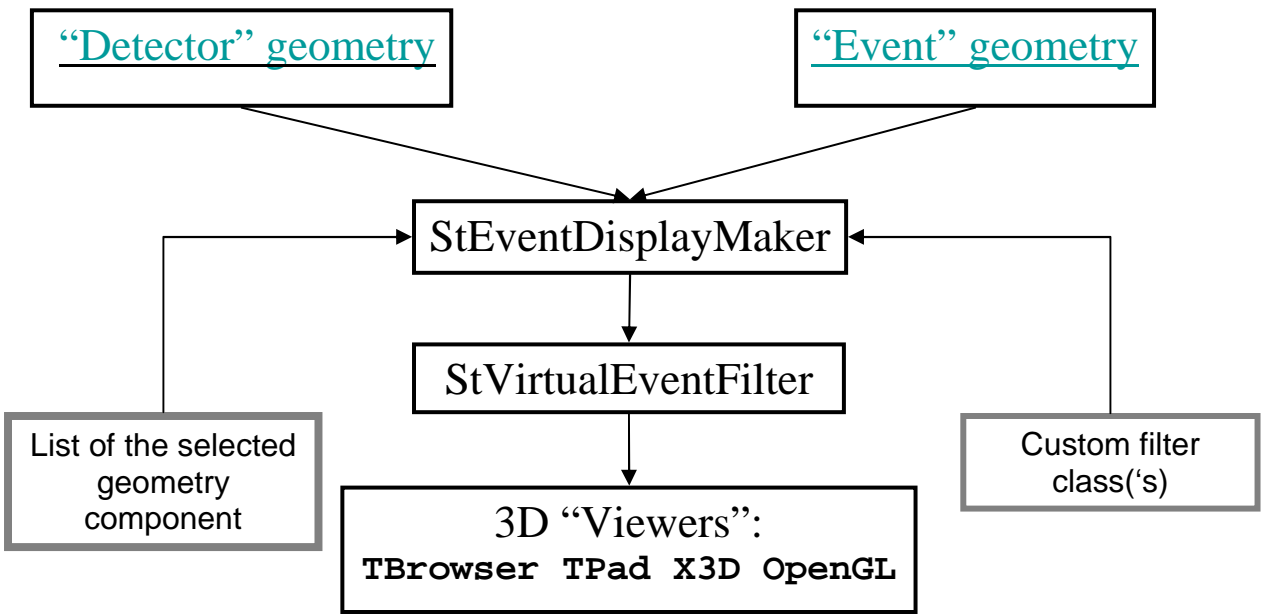

Figure 1: Object model

"Event Display" as sub-classes of the STAR base class StMaker [2] to be included in to STAR offline production chain [3] directly and communicate it via the advanced GUI interface customized with the Qt GUI Designer [4]

The object model of the 3D display includes the model of the so-called "Event Geometry", "Event Geometry' and "user event filter" [5] were implemented. (see Fig 1.)

## 2.2. Implementation

Event thought the model of the production Event Display was developed a while ago [2] there was not enough technical condition to implement it effectively and delivery the solution to the desktop of each researcher involved in the data-mining during the HEP experiment. The cheap and effective OpenGL PC video card, high resolution monitor, extremely fast CPU, and robust GUI cross platform libraries.

GUI of the "Display" consists of the 3 main panels:

1. 3D panel to visualize the detector and event geometry (Figure 2).

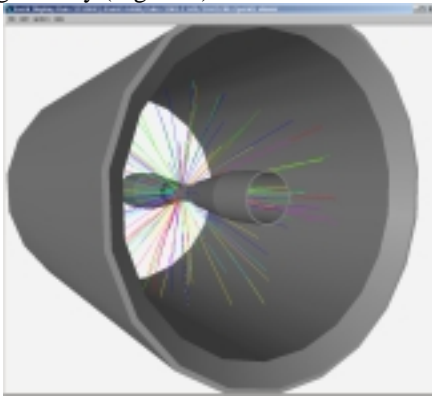

Figure 2: OpenGL Widget tp represent "Event" and "Detector" geometries after filtering.

2. Control Panel to provide the various component to allow user the customize the view and select the needed components ( Figure 3)

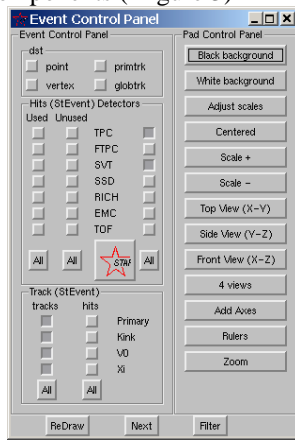

Figure 3: The Control Panel of the STAR Event Display.

3. The Filter panel. GUI of the user-provided filter C++ classes is generated automatically. The user should provide the selection bollena expression written in C++ language only ( Figure 4) .

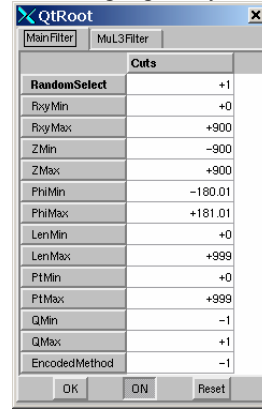

Figure 4: The example of the automatically created GUI for the user-provided filter class.

## 2.3. The dependency the "Event Display" package of STAR Framework

To make the developed software useful outside of the STAR collaboration all effort was done to minimize the package dependency of the STAR framework and avoid any connection with some custom file format.

The "StEventDisplayMaker" is a C++ class to combine the different "geometry" components. In addition the components can be passed through the chain of the custom data-filters.

This implemented schema is a rather generic. It contains the object of the ROOT C++ classes to represent the "Detector geometry" derived from the TShape ROOT class. To represent the components of the event geometry 3 based models are used, namely, the "segment" model to represent the portion of the tracks, the "dot" model to visualize the hits-like object and the "helix" model to draw the tracks.

## 3. CONCLUSION

During the job
- The software tools, OO model and C++ class library with the only ROOT dependency were developed to construct 3D "event display" widgets.
- That can be used to create custom displays adapted to the current reconstruction debugging needs.
- Real working event display was done and deployed as a part of the STAR reconstruction chain. That proved the principals are viable and the OO model is robust, the tools are easy to understand and customize.